\documentclass[12pt,a4paper]{article}

\usepackage{axodraw4j}
\usepackage{pstricks}
\usepackage{color}
\usepackage{epsfig}

\begin{document}
\def \lan {\langle}
\def \ran {\rangle}
\def \be  {\begin{equation}}
\def \ee  {\end{equation}}
\def \beq  {\begin{equation}}
\def \eeq  {\end{equation}}
\def \ba  {\begin{eqnarray}}
\def \ea  {\end{eqnarray}}
\def \baa {\begin{eqnarray*}}
\def \eaa {\end{eqnarray*}}
\def \lab #1 {\label{#1}}
\newcommand\bqa {\begin{eqnarray}}
\newcommand\eqa {\end{eqnarray}}
\newcommand\pr {\partial}
\newcommand\apr {\overline {\partial }}
\newcommand\nn {\nonumber}
\newcommand \noi {\noindent}
\newcommand{\bear}{\begin{array}}
\newcommand{\enar}{\end{array}}
\newcommand{\hf}{\frac{1}{2}}
\newcommand{\vx}{\vec{x}}
\newcommand{\R}{\mathbb{R}}
\newcommand{\C}{\mathbb{C}}
\newcommand{\Q}{\mathbb{Q}}
\newcommand{\F}{\mathbb{F}}
\newcommand{\A}{\overline{\mathbb{C}}}
\newcommand{\Z}{\mathbb{Z}}
\newcommand{\bg}{{\bf g}}
\newcommand{\Tpl}{{T}_+}
\newcommand{\Tmin}{\mathcal{T}_-}
\newcommand{\LL}{{L}}
\newcommand{\AAA}{\overline{A}}
\newcommand{\inv}[1]{{#1}^{-1}} 

\def\t{\theta}
\def\T{\Theta}
\def\w{\omega}
\def\ov{\overline}
\def\a{\alpha}
\def\b{\beta}
\def\g{\gamma}
\def\s{\sigma}
\def\l{\lambda}
\def\wt{\widetilde}

\def \CO {{\cal O}}
\def \CP {{\cal P}}
\def \CT {{\cal T}}
\def \CM {{\cal M}}
\def \CK {{\cal K}}
\def \CH {{\cal H}}
\def \CI {{\cal I}}
\def \CV {{\cal V}}
\def \CJ {{\cal J}}
\def \CL {{\cal L}}

\font\cmss=cmss12 \font\cmsss=cmss10 at 11pt
\def\inbar{\,\vrule height1.5ex width.4pt depth0pt}
\def\IC{\relax\hbox{$\inbar\kern-.3em{\rm C}$}}
\def\IZ{\relax{\hbox{\cmss Z\kern-.4em Z}}}
\def\IR{{\hbox{{\rm I}\kern-.2em\hbox{\rm R}}}}
\def\R{{\tiny \IR}}
\def\IP{{\hbox{{\rm I}\kern-.2em\hbox{\rm P}}}}
\def\II{\hbox{{1}\kern-.25em\hbox{l}}}

\renewcommand{\thefootnote}{\fnsymbol{footnote}}
\setcounter{footnote}{0}

\begin{titlepage}

\hfill\parbox{40mm}
{\begin{flushleft}  ITEP-TH-36/09
PUPT-2320
\end{flushleft}}

\vspace{10mm}
\centerline{\large \bf Aspects of
the ${\cal N}=4$ SYM amplitude - Wilson polygon duality}
\vspace{17mm}

\centerline{\bf A. Gorsky$^{1}$ and A. Zhiboedov$^{2,*}$\footnotetext{* On leave of absence from JINR BLTP Russia, Dubna}}

\vspace{10mm}
\begin{center}
{\it $^1$ Institute of Theoretical and Experimental Physics, Moscow, Russia, \\
$^2$Department of Physics, Princeton University, Princeton, NJ 08544, USA.
}
\vspace{1cm}
\end{center}

\vspace{1cm}

\vspace{1cm}

\centerline{\bf Abstract}
We discuss formulation of Wilson polygon - MHV amplitude duality
at the perturbative level in various regularizations. For four gluons it is shown that at one loop
one can formulate diagrammatic correspondence
interpolating  between the dimensional regularization
and the off-shell one.
We suggest new interpretation of all types of box diagrams in terms of the
dual simplex in dimensional regularization  and describe its degeneration
to the Wilson polygon. The interesting nullification phenomena for the
low-energy amplitudes in the Higgsed phase has been found.

\end{titlepage}

\section{Introduction}

One of the most surprising recent developments was the discovery of the
correspondence between the Wilson polygons
\begin{eqnarray} \label{wilsondef}
W({\cal C}_{n}) := \frac{1}{N} Tr {\cal P} \exp [ i g \oint d \tau  \dot{x}^{\mu}(\tau) A_{\mu} (x(\tau))]
\end{eqnarray}
and MHV scattering amplitudes in the ${\cal N}=4$ SYM
\beq
\mbox{Fin}\log[\frac{{\cal A}^{MHV}_{all-loop}}{{\cal A}^{MHV}_{tree}}]=\mbox{Fin}\log[\langle W(p_1,p_2,...,p_N) \rangle].
\label{am}
\eeq
It was first suggested at strong coupling \cite{am}
and later was clarified perturbatively at one-loop \cite{kor1, bt}, two-loop level
\cite{six1,six2} and at strong coupling  \cite{Berkovits:2008ic}. Since the origin of the correspondence was miraculous it is
necessary to make steps towards its clarification. In
\cite{Gorsky:2009nv} the derivation of one-loop MHV - Wilson loop duality was given
from the first principle.
The analysis  of   \cite{Gorsky:2009nv} clearly demonstrates that Feynman parameters in the loop integral
parameterize the space where the Wilson polygon is defined in.

In this paper we continue to investigate the explicit mapping between
two objects and shall focus on the several topics.
First, we make continuation from dimensional to
off-shell regularization for one-loop four gluons case and
formulate the correspondence.
Secondly, we consider Higgsed phase of ${\cal N}=4$ SYM recently considered in \cite{alday2},
where the VEV of scalars provide the regularization in the gauge invariant manner
and show that ``$AdS$'' regularized Wilson loop - Higgsed ${\cal N}=4$ SYM amplitude duality,
while works at one loop breaks down at two loops.
Third, we shall consider one loop box diagrams harder than 2me boxes. These  diagrams
enter the answers for non-MHV one loop amplitudes in dimensional regularization
\beq
{\cal A}_{n;1} =i (2 \pi)^4 \delta^{4}(p) \sum \left( {\cal C}^{4m} I^{4m} + {\cal C}^{3m} I^{3m} + {\cal C}^{2mh} I^{2mh}+{\cal C}^{2me} I^{2me}+{\cal C}^{1m} I^{1m}\right).
\eeq
We shall demonstrate that in this case there exists dual object, namely, simplex which
get reduced to the  Wilson polygon for MHV amplitude.

The  object dual to boxes harder than two-mass easy which appear in N$^{k}$MHV  amplitudes
turns out to be not polygon, but a simplex. The origin of the simplex is fixed by claiming
the dual edges to be light-like.
Namely one has to add the additional vertex connected with the
vertices of the Wilson polygon and the gauge field propagator is attached
just to this extra vertex of the simplex.
We demonstrate  that at one loop level this picture degenerates to the known duality when
we reduce generic box down to the 2me one. In the case of triangle diagram
we show that in the proper limit  the dual simplex for the 3m diagram reduces to the Wilson
triangle with the additional vertex found in \cite{Gorsky:2009nv}.

In the last part we will discuss  how one can see $AdS$ geometry
from weak coupling point of view from
both amplitude and Wilson loop side.  In the Higgsed phase we also provide the recipe
of the calculation of the low-energy non-MHV amplitude from the effective
actions in the external field and observe the peculiar properties of the amplitudes.

The paper is organized as follows. In Section 2 we derive
the duality to the ``off-shell'' regime at one loop level via the explicit
mass dependent change of variables.
Section 3 concerns the interpretation
of the 2mh and harder boxes in terms of dual simplex.
Section 4 is devoted
to some two-loop dual calculation in the ``$AdS$'' regularization.
In section 5 we show how $AdS_3$ geometry emerges from massless
amplitudes and Wilson loop calculations and
in Section 6 the low-energy amplitudes in the Higgsed phase have been obtained.
In the last Section we shall make comments on the open questions. In Appendices
we collect some relevant notations and integrals as well as present an interesting
geometrical interpretation behind the divergencies of the integrals.

\section{Going off-shell from four gluons amplitude - Wilson polygon one-loop duality}

Here we discuss the duality for four gluon amplitude at one loop and its possible extension
off-shell generalizing the  approach of \cite{Gorsky:2009nv} to the case
of off-shell external particles. In the following section $D_{IR}=4 + 2 \epsilon$ and
$D_{UV}=4 - 2 \epsilon$.

\subsection{On-shell form of duality}
Let us remind the formulation of the on-shell duality.
At one loop the vacuum expectation of light-like Wilson loop has the following structure
\begin{eqnarray}
{\cal W} = 1 -  a  (\pi \mu_{UV}^2)^{\epsilon} \Gamma(1 - \epsilon) \sum_{1 \leq j,k \leq 4} \tilde{I}^{W}_{j k}(4 - 2 \epsilon)
\end{eqnarray}
where we have six $(jk)$ diagrams $(12,13,14,23,24,34)$, or three pairs of different diagrams. Notations which we use
are given in Appendix A .

Four-gluon amplitude at one loop is given by the following expression
\begin{eqnarray}
{\cal M}_{4}^{MHV} =1 -\frac{a}{2}  s u I_{4}^{0m}(4 + 2 \epsilon) =1 - a (\mu_{IR}^2)^{\epsilon} \Gamma(1 - \epsilon) \frac{\Gamma(1 + \epsilon)^2}{\Gamma(1 + 2 \epsilon)} F_{4}(4 + 2 \epsilon)
\end{eqnarray}
where $I_{4}^{0m}$ is massless box diagram.

In this particular case duality can be stated in  several different forms:

\textbf{Version 1 (stronger one).}
\begin{eqnarray}
\sum_{1 \leq j,k \leq 4} \tilde{I}^{W}_{j k} (4 - 2 \epsilon) = F_{4}(4 + 2 \epsilon)
\end{eqnarray}
which is true up to all orders in $\epsilon$. Such form of the duality is known to be violated
at higher loops due to different divergent sub-leading parts which are governed by different
functions.
This version also can be rewritten as the following equality
\begin{eqnarray}
\frac{\Gamma(1 + \epsilon)^2}{\Gamma(1 + 2 \epsilon)} {\cal W} = {\cal M}_{4}
\end{eqnarray}

\textbf{Version 2 (weaker one).}
\begin{eqnarray}
\mbox{Fin}\log[ {\cal W}] =\mbox{Fin}\log[{\cal M}_{4}].
\end{eqnarray}
where $\mbox{Fin}[...]$ means equality of finite parts up to arbitrary kinematics independent constant.
Up to date the most powerful check of this statement is two-loop six gluons calculation \cite{six1,six2}.
Here we will show how the duality in the strong form can be pushed off-shell and what terms can be
attributed to each Wilson diagram in the dual picture.

\subsection{Connection between massive box diagrams in different dimensions}
Remind that in our derivation \cite{Gorsky:2009nv} of the duality in the dimensional regularization two step were important. First, the peculiar change of variables in the space of the Feynman parameters  and secondly the relation between the box diagrams in $D$ and $(D-2)$ dimensions \cite{Tarasov}.
To formulate the duality in the "off-shell" case
let us  use once again the connection between integrals in different dimensions that is given in the appendix A for the
box diagram with $D=6+2\epsilon$ and $p_{i}^2 = m^2$ .
For this case we get
\begin{eqnarray}
I_{4}(6 + 2 \epsilon, m) = \frac{1}{(1 + 2 \epsilon) z_{0}} \left(I_{4}(4 + 2 \epsilon, m) - \sum_{i=1}^{4} z_{i} I_{3}(4 + 2 \epsilon, m ; 1 - \delta_{k i} ) \right)
\end{eqnarray}
where
\begin{eqnarray}
z_{0} = \sum_{i=1}^{4} z_{i} &=& 2 \frac{s + u - 4 m^{2}}{s u - 4 m^4} \\ \nonumber
z_{1} &=& z_{3} = \frac{u- 2 m^{2}}{s u - 4 m^4} \\ \nonumber
z_{2} &=& z_{4} = \frac{s- 2 m^{2}}{s u - 4 m^4} \\ \nonumber
\end{eqnarray}
Let's rewrite equality between integrals in the following useful way
\begin{eqnarray}\label{duality}
&-\frac{a}{2} s u I_{4}(4 + 2 \epsilon, m) = a \frac{\frac{\Gamma(1 + \epsilon)^2}{\Gamma(1 + 2 \epsilon)}}{1 -  \frac{4 m^4}{s u}} \\ \nonumber
&[ 2 w_{0} \frac{\Gamma(2 + 2 \epsilon)}{\Gamma(1 +\epsilon)^2} I_{4}(6 + 2 \epsilon, m) - \sum_{i=1}^{4} w_{i} \frac{\Gamma(1 + 2 \epsilon)}{\Gamma(1 + \epsilon)^2} I_{3}(4 + 2 \epsilon, m ; 1 - \delta_{k i} )]
\end{eqnarray}
\begin{eqnarray}
w_{0} &=& \frac{4 m^{2} - s - u}{2} = (p_{1},p_{3}) = (p_{2},p_{4})\\ \nonumber
w_{1} &=& \frac{u - 2 m^{2}}{2} = (p_{2},p_{3})\\ \nonumber
w_{2} &=& \frac{s - 2 m^{2}}{2} = (p_{3},p_{4})\\ \nonumber
w_{3} &=& \frac{u - 2 m^{2}}{2} = (p_{4},p_{1})\\ \nonumber
w_{4} &=& \frac{s - 2 m^{2}}{2} = (p_{1},p_{2})\\ \nonumber
\end{eqnarray}
On the other hand one-loop correction to the amplitude in ${\cal N}=4$ for four gluons is given by
\beq
{\cal M}_{4} = -\frac{a}{2}  s u I_{4}(4 + 2 \epsilon, m)|_{m \rightarrow 0}
\eeq
\beq
{\cal M}_{4}= - \frac{a}{2} s u I_{4}(4 + 2 \epsilon, m)|_{\epsilon \rightarrow 0}
\eeq
in dimensional and mass regularizations.
If  $m=0$  each summand in the RHS of (\ref{duality}) becomes equal exactly to the corresponding Wilson diagram and we get
\begin{eqnarray}
{\cal M}_{4} = \frac{\Gamma(1 + \epsilon)^2}{\Gamma(1 + 2 \epsilon)} {\cal W}
\end{eqnarray}
that is our strong version of the duality mentioned above.
Thus, (\ref{duality}) can be considered as the continuation of the stronger version of the duality off-shell.


Now we can  send $\epsilon \rightarrow 0$,
and get the analog of each Wilson diagram in this case.
We have the following changes (strictly speaking since we have no strict definition of off-shell ``Wilson loop'', we are free to multiply and divide by the function $f(m)$ which is $f(0)=1$ to redefine off-shell diagrams)

\begin{eqnarray}
\frac{\Gamma(1 + \epsilon)^2}{\Gamma(1 + 2 \epsilon)} {\cal W} = {\cal M}_{4} &\leftrightarrow& \frac{1}{1 -  \frac{4 m^4}{s u}}  {\cal W} = {\cal M}_{4} \\
\frac{I^{cusp}_{i i+1}(4 - 2 \epsilon|s_{i,i+1})}{(p_{i},p_{i+1})} &\leftrightarrow& - I_{3}(4|s_{i,i+1},m^2,m^2) \\
\frac{I^{W}_{i j}(4 - 2 \epsilon|s,u)}{(p_{i},p_{j})} &\leftrightarrow& I_{4}(6|s,u,m^2,m^2,m^2,m^2)
\end{eqnarray}
There is no duality with the Wilson polygon in this case since objects at the r.h.s. have no such interpretation. In the next Section we suggest the proper geometrical object which substitutes
Wilson polygons.

Note that cusp diagram in this case is equal to
\begin{eqnarray}
I^{cusp}_{i i+1}(4 |s_{i,i+1}) = -  \frac{1}{2} \log^2 (\frac{-m^2}{s_{i,i+1}})
\end{eqnarray}
Here the coefficient is important, because the amplitude
in the off-shell regularization satisfies the equation
\beq
(\frac{\partial}{\partial \ln (m^2)})^2 \ln M_4 = - 2 \Gamma_{cusp}(a)
\eeq
with extra factor of $2$. Up to vanishing in $m \to 0$ terms this result is reproduced
via cutting prescription discussed in \cite{Dorn:2008dz}.

\begin{figure}
\begin{center}
\fcolorbox{white}{white}{
  \begin{picture}(212,203) (127,-79)
    \SetWidth{1.0}
    \SetColor{Black}
    \Line(128,-27)(203,-77)
    \Text(140,-53)[lb]{\Black{$p_{1}$}}
    \Line(203,-78)(303,-40)
    \Line(291,10)(128,-28)
    \Line[arrow,arrowpos=0.5,arrowlength=5,arrowwidth=2,arrowinset=0.2](178,123)(203,-78)
    \Gluon(178,123)(241,-40){3.5}{15}
    \Text(253,-72)[lb]{\Black{$p_{2}$}}
    \Text(178,-12)[lb]{\Black{$p_{4}$}}
    \Text(203,-40)[lb]{\Black{$q_{2}$}}
    \Line(304,-40)(291,10)
    \Line[arrow,arrowpos=0.5,arrowlength=5,arrowwidth=2,arrowinset=0.2](178,123)(128,-28)
    \Line[arrow,arrowpos=0.5,arrowlength=5,arrowwidth=2,arrowinset=0.2](178,123)(304,-40)
    \Line[arrow,arrowpos=0.5,arrowlength=5,arrowwidth=2,arrowinset=0.2](178,123)(291,10)
    \Text(304,-15)[lb]{\Black{$p_{3}$}}
    \Text(140,48)[lb]{\Black{$q_{1}$}}
    \Text(238,73)[lb]{\Black{$q_{4}$}}
    \Text(238,23)[lb]{\Black{$q_{3}$}}
    \SetColor{Red}
    \Vertex(241,-40){3}
    \SetColor{Black}
    \Vertex(178,123){3}
  \end{picture}
}
\end{center}
\caption{\small Basic simplex in dual space.}
\end{figure}
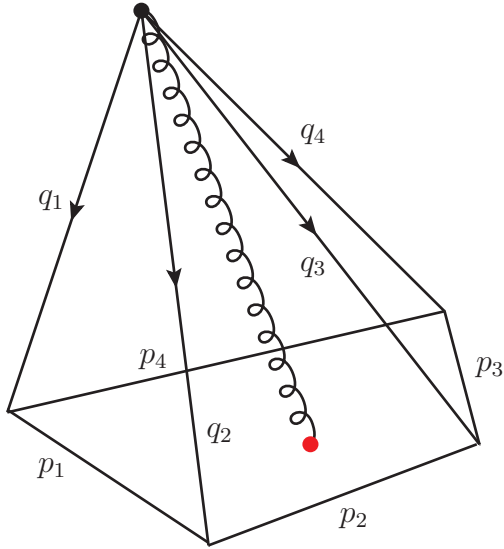

\section{Towards the possible interpretation of arbitrary box in dual  terms}

In the massless case the amplitude is expected to be dual to the Wilson
polygon however the generalization to the massive case is nor immediate.
In this Section we shall consider the object dual to the
arbitrary massive box diagrams and it turns out that the polygon has to be
substituted by more generic simplex.

To start with note that the arbitrary box in Feynman parametrization is defined in terms of the following integral
\begin{eqnarray}
I &=& \int \prod d x_{i} \frac{\delta( 1 - x_1 - x_2 - x_3 - x_4)}{(-\Delta)^{4 - \frac{D_{IR}}{2}}} \\ \nonumber
\Delta &=& s x_1 x_3+ u x_2 x_4 +m_{1}^2 x_{1} x_{2} + m_{2}^2 x_{2} x_{3} + m_{3}^2 x_{3} x_{4}+ m_{4}^2 x_{4} x_{1} \\ \nonumber
\end{eqnarray}
where $s=(p_1+p_2)^2=(p_3+p_4)^2$ and $u=(p_2+p_3)^2=(p_1+p_4)^2$.
Let's introduce dual coordinates $p_{i} = q_{i} - q_{i+1}$ (we use $q$ instead of $x$ to avoid mixing with Feynman parameters).
Then one can rewrite
\begin{eqnarray}
\Delta &=& q_{1}^2 x_{1}+q_{2}^2 x_{2}+q_{3}^2 x_{3}+q_{4}^2 x_{4}-(q_1 x_{1} + q_{2} x_{2} + q_{3} x_{3} + q_{4} x_{4})^2 \nonumber
\end{eqnarray}
and assuming that dual vectors are light-like (or using dual translational invariance),
\beq
\Delta =-(q_1 x_{1} + q_{2} x_{2} + q_{3} x_{3} + q_{4} x_{4})^2.
\eeq
This form suggests that it could be interpreted as massless propagator in position space.
Note that condition $q_{i}^2 = 0$  can not be imposed in general kinematics.
Now the box diagram can be interpreted in terms of the object which we will call
following \cite{dav} \emph{basic simplex} (see fig.1).

\begin{figure}
\begin{center}
\fcolorbox{white}{white}{
  \begin{picture}(228,174) (76,-30)
    \SetWidth{1.0}
    \SetColor{Black}
    \Line(211,-29)(132,24)
    \Line(132,24)(211,143)
    \Line(211,142)(303,50)
    \Line(303,50)(211,-29)
    \Line[dash,dashsize=10](132,24)(303,50)
    \Line(211,142)(211,-29)
    \Text(145,-16)[lb]{\Large{\Black{$p_{1}$}}}
    \Text(132,89)[lb]{\Large{\Black{$p_{2}$}}}
    \Text(264,103)[lb]{\Large{\Black{$p_{3}$}}}
    \Text(264,-3)[lb]{\Large{\Black{$p_{4}$}}}
    \Vertex(79,37){3}
    \Gluon(79,37)(184,63){3}{9}
    \SetColor{Red}
    \Vertex(185,63){3}
  \end{picture}
}
\end{center}
\caption{\small The region of integration.}
\end{figure}
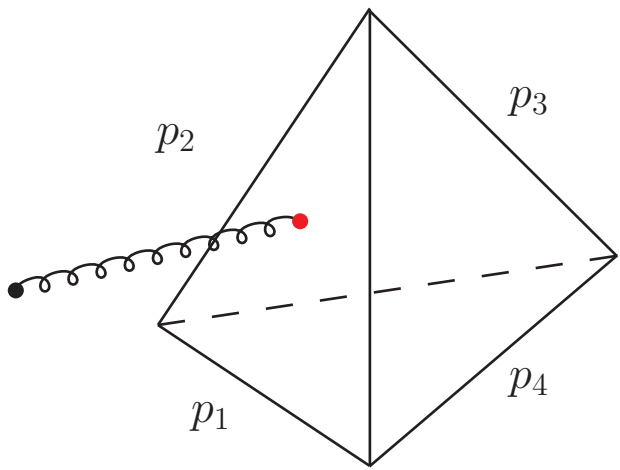
One should notice that general $n$-point diagram with massive propagators will be represented by
\begin{eqnarray}
\Delta_{n} &=& \sum (q_{i}^2 - m_{i}^2) x_{i} -(\sum q_i x_{i} )^2 \nonumber
\end{eqnarray}
where $m_{i}$ are propagator masses.

Feynman parameters delta-function bounds the integration region to the momenta tetrahedron for which momenta $p_{i}^{\mu}$, $s^{\mu}=(p_{1}+p_{2})^{\mu}$ and $u^{\mu}=(p_{2}+p_{3})^{\mu}$ serves as edges (see fig.2).
The suggested picture reproduces the known picture for the
two-mass easy case $p_{1}^2=0=p_{3}^2$. To get this situation we can choose $q_{1}||q_{2}$ and $q_3 || q_{4}$ (see fig.3).

Taking into account that $q_{2} = q_1 - p_1$ and $q_4 = q_3 - p_3$ we can obtain
\begin{eqnarray}
\Delta^{2me} &=&-(q_1 (x_{1}+x_{2}) - p_{1} x_{2} + q_{3} (x_{3} + x_{4}) - p_{4} x_{4})^2 \\ \nonumber
&=& -((x_{1}+x_{2})(q_1 - p_1 \frac{x_2}{x_{1}+x_{2}})+(x_3+x_4)(q_{3} - p_{4} \frac{x_{4}}{x_3+x_4}))^2
\end{eqnarray}
Let us introduce the following variables
\beq
\tau_1 = \frac{x_2}{x_{1}+x_{2}}\qquad  \tau_2 = \frac{x_{4}}{x_3+x_4}
\eeq.
It is evident  that both
$q_1 + p_1 \frac{x_2}{x_{1}+x_{2}}$ and $q_{3} + p_{4} \frac{x_{4}}{x_3+x_4}$ are light-like hence
we can write $$\Delta^{2me}= (x_{1}+x_{2})(x_{3}+x_{4}) \tilde{\Delta}(\tau_1,\tau_2).$$
where $\tilde{\Delta}(\tau_1,\tau_2)$ is nothing but Wilson loop diagram propagator (see fig. 4). It turns out that the measure is also reproduced
in the right way.

\begin{figure}
\begin{center}
\fcolorbox{white}{white}{
  \begin{picture}(267,145) (60,-123)
    \SetWidth{1.0}
    \SetColor{Red}
    \Vertex(198,-119){3.162}
    \SetColor{Black}
    \Line[double,sep=2](197,-118)(62,-1)
    \Line[double,sep=2](198,-118)(314,-17)
    \Vertex(64,-2){3.162}
    \Vertex(155,-81){3.162}
    \Vertex(266,-61){3.162}
    \Vertex(313,-18){3.162}
    \Line[arrow,arrowpos=0.5,arrowlength=5,arrowwidth=2,arrowinset=0.2](157,-82)(265,-61)
    \Line[arrow,arrowpos=0.5,arrowlength=5,arrowwidth=2,arrowinset=0.2](312,-19)(66,-3)
    \Text(87,-48)[lb]{\Large{\Black{$p_1$}}}
    \Text(292,-48)[lb]{\Large{\Black{$p_3$}}}
    \Text(179,1)[lb]{\Large{\Black{$p_4$}}}
    \Text(192,-64)[lb]{\Large{\Black{$p_2$}}}
    \Text(157,-110)[lb]{\Large{\Black{$q_1$}}}
    \Text(244,-96)[lb]{\Large{\Black{$q_3$}}}
    \Gluon(197,-119)(164,-39){3}{6}
    \SetColor{Red}
    \Vertex(165,-39){3.162}
  \end{picture}
}
\end{center}
\caption{\small Degeneration of basic simplex in the case
of two-mass easy box diagram.}
\end{figure}
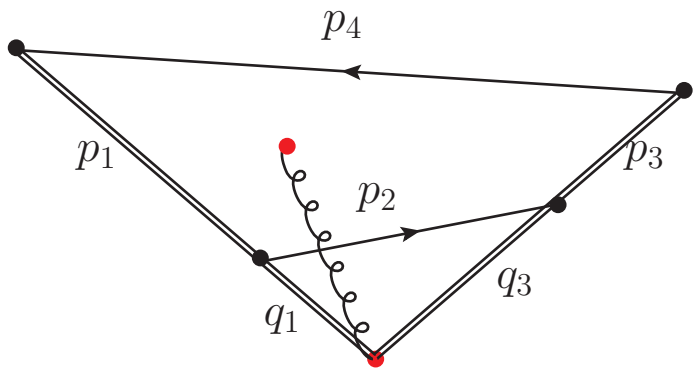

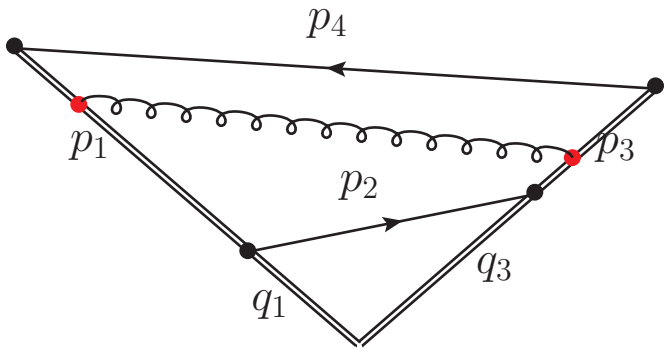
\begin{figure}
\begin{center}
\fcolorbox{white}{white}{
  \begin{picture}(257,138) (60,-123)
    \SetWidth{1.0}
    \SetColor{Black}
    \Vertex(258,-64){3.162}
    \Vertex(303,-24){3.162}
    \Line[double,sep=2](192,-121)(63,-9)
    \Line[double,sep=2](193,-121)(304,-24)
    \Vertex(64,-9){3.162}
    \Vertex(151,-86){3.162}
    \Line[arrow,arrowpos=0.5,arrowlength=5,arrowwidth=2,arrowinset=0.2](153,-86)(257,-65)
    \Line[arrow,arrowpos=0.5,arrowlength=5,arrowwidth=2,arrowinset=0.2](302,-25)(66,-10)
    \Text(85,-52)[lb]{\Large{\Black{$p_1$}}}
    \Text(282,-53)[lb]{\Large{\Black{$p_3$}}}
    \Text(174,-6)[lb]{\Large{\Black{$p_4$}}}
    \Text(186,-67)[lb]{\Large{\Black{$p_2$}}}
    \Text(153,-112)[lb]{\Large{\Black{$q_1$}}}
    \Text(237,-98)[lb]{\Large{\Black{$q_3$}}}
    \SetColor{Red}
    \Vertex(88,-31){3.162}
    \Vertex(272,-51){3.162}
    \SetColor{Black}
    \Gluon(89,-30)(272,-51){3}{13}
  \end{picture}
}
\end{center}
\caption{\small Appearance of Wilson loop diagram from the degenerate basic simplex. By double lines light-like edges are denoted. }
\end{figure}

\subsection{3-point function case}

In the case of 3-point function situation is quite similar. The only
difference is that instead of four vectors from the origin of the simplex we have three ones.

To check the suggested picture let us reproduce the known picture for the
two-mass hard case $p_{1}^2=0$. To get this situation we can choose $q_{1}||q_{2}$.
Let's use the fact that $q_{2} = q_1 - p_1$ to rewrite
\begin{eqnarray}
\Delta^{2mh} &=&-(q_1 (x_{1}+x_{2}) - p_{1} x_{2} + q_{3} x_{3})^2 \\ \nonumber
&=& -((x_{1}+x_{2})(q_1 - p_1 \frac{x_2}{x_{1}+x_{2}})+ x_3 q_{3})^2
\end{eqnarray}
Introduce $\tau = \frac{x_2}{x_{1}+x_{2}}$ and note that both
$q_1 + p_1 \frac{x_2}{x_{1}+x_{2}}$ and $q_{3}$ are light-like. Therefore
we can write
\beq
\Delta^{2mh}= (x_{1}+x_{2}) x_{3} \tilde{\Delta}(\tau).
\eeq
where $\tilde{\Delta}(\tau)$ is Wilson loop diagram propagator (see fig. 5).

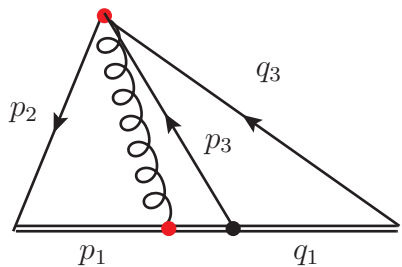
\begin{figure}
\begin{center}
\fcolorbox{white}{white}{
  \begin{picture}(153,96) (72,-76)
    \SetWidth{1.0}
    \SetColor{Black}
    \Line[double,sep=2](223,-64)(81,-64)
    \Line[arrow,arrowpos=0.5,arrowlength=5,arrowwidth=2,arrowinset=0.2](224,-63)(113,16)
    \Line[arrow,arrowpos=0.5,arrowlength=5,arrowwidth=2,arrowinset=0.2](113,16)(80,-65)
    \Vertex(162,-64){2.828}
    \Gluon(113,15)(136,-63){4.5}{7}
    \SetColor{Red}
    \Vertex(138,-64){2.828}
    \Vertex(114,16){2.828}
    \Text(171,-9)[lb]{\Black{$q_{3}$}}
    \Text(185,-78)[lb]{\Black{$q_{1}$}}
    \Text(105,-78)[lb]{\Black{$p_{1}$}}
    \SetColor{Black}
    \Line[arrow,arrowpos=0.5,arrowlength=5,arrowwidth=2,arrowinset=0.2,flip](114,17)(163,-65)
    \Text(79,-24)[lb]{\Black{$p_{2}$}}
    \Text(152,-36)[lb]{\Black{$p_{3}$}}
  \end{picture}
}
\end{center}
\caption{Dual to two-mass triangle.}
\end{figure}

To conclude this section let us briefly comment on the possible strong
coupling counterpart of our weak coupling picture. The natural candidate
for the massive gluon amplitude is the correlator of the Wilson polygon with the
local operator inserted into the vertex of the simplex $\langle W({\cal C})O \rangle$. Such
correlators were investigated in the duality context, for instance in \cite{malda}
and at strong coupling it reduces to the exchange by some string mode.
We have presented only one-loop arguments supporting the correct choice of the dual
object for the off-shell external legs.
However it is clear that more detailed analysis is desired including two-loop calculations.

\section{``$AdS$'' regularized Wilson loop - Higgsed regularized amplitude duality at two loops}
In this Section we shall try the different regularization scheme to discuss the Higgsed phase
of the theory. Our aim is to suit the proper regularization for the discussion of two-loop off-shell duality.
In particular we will show that duality between  naively ``$AdS$'' regularized Wilson loop and Higgsed amplitude
is valid at one loop but breaks down at two loops. We use the fact of exponentiation recently found in \cite{alday2}
at two loops and explicitly show that $\ln^4 m^2$ do not cancel in Wilson loop calculation, while such terms cancel
on the amplitude side of the correspondence.


The perturbative expansion of Wilson loop
\beq
\label{wil}
W[ {\cal C}_n]  \ := \frac{1}{N}\ {\rm Tr} \, {\cal P} \exp \left[ i g\oint_{{\cal C}_n} \! d\tau  \ \dot{x}^{\mu} (\tau )A_\mu (x(\tau ))   \right]
\ .
\eeq
reads as follows
\ba
\label{naeA}
\lan W[{\cal C}_n ] \ran   &=&  1 \, + \, \sum_{l=1}^{\infty} a^l W^{(l)}_n \ = \   \exp \sum_{l=1}^{\infty} a^l w^{(l)}_n \ , \\
\label{w2ourA}
w^{(2)}_n &=&  W^{(2)}_n \, - \, {1\over2} \, (W^{(1)}_n)^2
\ ,
\ea
where $a = \frac{g^2 N}{8 \pi^2}$ ( $C_{F}=\frac{N}{2}$ and $C_{A}=N$ in the planar limit).

Here we introduce - as a prescription - $AdS$ regularized analogue of Feynman
gauge propagator $\Delta_{\mu \nu} (x) := \eta_{\mu \nu} \Delta(x)$, where
\beq
\Delta  (x)  =
 {1\over 4 \pi^2}
 {1\over x^2 + z_{UV}^2- i \epsilon }
\eeq
and in the spirit of duality with amplitudes we will try to
identify $z_{UV}$ with regulator mass $m$ in the
Higgsed $U(N) \times U(n) $ model.

Three-point vertex is given by
\beq
-i C_F f^{a_1 a_2 a_3} \times
 \ \Big[ \eta^{\mu_1 \mu_2} (\partial^{\mu_3}_1 - \partial^{\mu_3}_2 )  +
\eta^{\mu_2 \mu_3} (\partial^{\mu_1}_1 - \partial^{\mu_1}_2 ) + \eta^{\mu_3 \mu_1} (\partial^{\mu_2}_1 - \partial^{\mu_2}_2 )
\Big] G(x_1, x_2, x_3)
\ ,
\eeq
where $G(x_1, x_2, x_3) \ = \ \int\!\! d^Dz \ \Delta (x_1 -z) \Delta (x_2 -z) \Delta (x_3 -z)$
\beq \label{G}
G(x_1, x_2, x_3) \ = \ {i \over 64 \pi^4 } \int\!\!\prod_{i=1}^{3} d \a_i \ \delta(1 - \sum_{i=1}^3 \a_1 )
{1 \over \a_1 \a_2 x_{12}^2 + \a_1 \a_3 x_{13}^2 + \a_2 \a_3 x_{23}^2 + m^2} \ , \\ \nonumber
\eeq
and we have used $x_{ij}^2 = (x_i - x_j)^2$.

To verify or disprove the duality and identification
of the regularization scales at one loop
we need calculate here only cusp diagrams since finite diagrams
do not depend on regularization.
For cusp diagram (see fig.6) we get
\begin{figure}
\begin{center}
\fcolorbox{white}{white}{
  \begin{picture}(148,124) (61,-71)
    \SetWidth{1.0}
    \SetColor{Black}
    \Line[double,sep=2](111,-69)(63,-21)
    \Line[double,sep=2](63,-21)(111,51)
    \Line[double,sep=2](111,51)(206,3)
    \Line[double,sep=2](207,3)(112,-69)
    \GluonArc(81,-15)(30.594,-78.69,78.69){3}{6}
  \end{picture}
}
\end{center}
\caption{\small Cusp diagram. }
\end{figure}
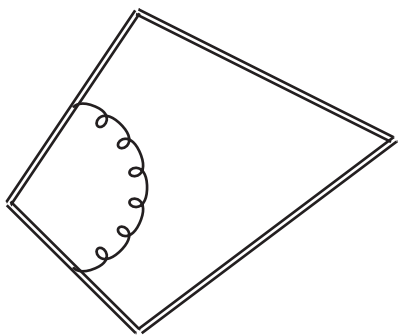
\beq
w_{i,i+1} = \frac{x_{i,i+1}^2}{2} \int_{0}^{1} d \tau_1 d \tau_2 \frac{1}{- x_{i,i+1}^2 (1 - \tau_1) \tau_2 - m^2}
\eeq
and  simple Mellin-Barnes technics yields
\beq
w_{i,i+1} = - \frac{1}{4} \ln^2 (\frac{x_{i,i+1}^2}{m^2})
\eeq
\beq
w_4^{(1)cusps} = - \frac{1}{2} (\ln^2 (\frac{s^2}{m^2}) + \ln^2 (\frac{u^2}{m^2}))
\eeq
Using analogy with dimensional regularization we can define
\beq
(\frac{\partial}{\partial \ln (m^2)})^2 \ln W_4 = - \Gamma_{cusp}(a)
\eeq
To first order it properly reproduces the known answer
\beq
\Gamma_{cusp}(a) = 2 a
\eeq


At two loops we will concentrate at $\ln^4 m^2$
terms and will show that they do not cancel  contrary
to the amplitude side. Thus, using notations of \cite{alday2}
even in $m_{i} = m$ regime the naive $AdS$ regularization
does not do the job. For the notations and general expressions
in dimensional regularization see \cite{Anastasiou:2009kna}.

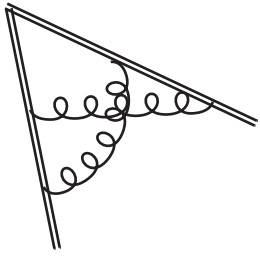
\begin{figure}
\begin{center}
\fcolorbox{white}{white}{
  \begin{picture}(97,94) (84,-55)
    \SetWidth{1.0}
    \SetColor{Black}
    \Line[double,sep=2](105,-54)(87,36)
    \Line[double,sep=2](179,-7)(86,37)
    \GluonArc(93.575,2.566)(34.176,-79.164,23.146){4}{5}
    \Gluon(163,1)(95,-2){4}{5}
  \end{picture}
}
\end{center}
\caption{\small Degenerate cross diagram. }
\end{figure}

The fig.7 diagram is given by
\beq
w_{d} = - (\frac{x_{i,i+1}^2}{2})^2 \int_{0}^{1} d \sigma_1 d \tau_1 \int_{0}^{\sigma_1} d \tau_2 \int_0^{\tau_2} d \sigma_2 \frac{1}{ x_{i,i+1}^2 \sigma_1 \sigma_2 + m^2} \frac{1}{x_{i,i+1}^2 \tau_1 \tau_2 + m^2}
\eeq
We introduce two-fold Mellin-Barnes representation and make all the $\tau$ and $\sigma$ integrations to get
\beq
w_{d} = - \frac{1}{4} \int d z_1 d z_2 (\frac{m^2}{x_{i,i+1}^2})^{z_1 + z_2} \frac{\Gamma(- z_1) \Gamma(- z_2) \Gamma(1+z_1) \Gamma(1+z_2)}{z_1 z_2 (z_1 + z_2)^2}
\eeq
The deformation  of the contour of integration amounts to $\ln^2 (m^2)$ terms which are unnecessary for our consideration
\beq
w_{d}^{AdS} = - \frac{1}{48} \ln^4 (\frac{x_{i,i+1}^2}{m^2})
\eeq

\beq
w_{d}^{dim.red.} = - \frac{1}{16} \frac{1}{\epsilon^4}
\eeq


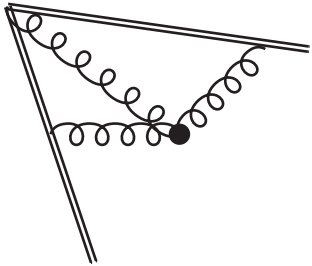
\begin{figure}
\begin{center}
\fcolorbox{white}{white}{
  \begin{picture}(115,99) (46,-62)
    \SetWidth{1.0}
    \SetColor{Black}
    \Line[double,sep=2](80,-61)(48,35)
    \Line[double,sep=2](160,19)(48,35)
    \Gluon(64,-13)(112,-13){4}{4}
    \Gluon(112,-13)(144,19){4}{4}
    \Gluon(112,-13)(48,35){4}{6}
    \Vertex(112,-13){4}
  \end{picture}
}
\end{center}
\caption{\small Y diagram. }
\end{figure}
For diagram depicted on fig. 8 we get\footnote{Remember that here we get factor of 2 from ``upside-down'' diagram}
\beq
w_{Y} = \frac{1}{2} \frac{x_{i,i+1}^2}{2} \int_{0}^{1} d \tau_1 d \tau_2 \int_0^1 \prod_1^3 d \alpha_{i} \frac{\delta( 1 - \alpha_1 - \alpha_2 - \alpha_3)}{x_{i,i+1}^2 \tau_1 \tau_2 \alpha_2 \alpha_3 + m^2}
\eeq

\beq
w_{Y}^{AdS} = \frac{1}{96} \ln^4 (\frac{x_{i,i+1}^2}{m^2})
\eeq

\beq
w_{Y}^{dim.red.} = \frac{1}{16} \frac{1}{\epsilon^4}
\eeq

Thus, we see that while cancel in dimensional reduction scheme these terms
do not cancel in $AdS$ regularization. Hence we have demonstrated that
naive $AdS$ inspired regularization of the propagator does not work
at two loop level.

\section{$AdS$ geometry from weak coupling}

It was shown in  \cite{gopa} that one can  recognize $AdS$ geometry starting from the
Feynman diagrams. Here we will show how $AdS$ geometry appears in the different  manner
\cite{dav} from the massless box diagram.


\begin{eqnarray}
I &=& \int_{0}^{\infty} d x_{1} d x_{2} d x_{3} d x_{4} \frac{\delta (1 - x_{1} - x_{2}- x_{3} - x_{4})}{(s x_{1} x_{3} + u x_{2} x_{4})^{4-\frac{D}{2}}}  \\ \nonumber
&=& \frac{1}{(- s u)^{2 - \frac{D}{4}}}\int_{0}^{\infty} d x_{1} d x_{2} d x_{3} d x_{4} \frac{\delta (1 - x_{1} - x_{2}- x_{3} - x_{4})}{(\sqrt{- \frac{s}{u}} x_{1} x_{3} + \sqrt{- \frac{u}{s}} x_{2} x_{4})^{4-\frac{D}{2}}} \\ \nonumber
\alpha^2 &=& - \frac{s}{u} > 0
\end{eqnarray}

Now let's make change of variables \cite{dav} $\sum y_{i} = y_{1}+y_{2}+y_{3}+y_{4}$

\begin{eqnarray}
x_1 &=& \frac{\alpha y_{1} y_{3} + \frac{1}{\alpha} y_{2} y_{4}}{\sum y_{i}} y_{1} \\ \nonumber
x_2 &=& \frac{\alpha y_{1} y_{3} + \frac{1}{\alpha} y_{2} y_{4}}{\sum y_{i}} y_{2} \\ \nonumber
x_3 &=& \frac{\alpha y_{1} y_{3} + \frac{1}{\alpha} y_{2} y_{4}}{\sum y_{i}} y_{3} \\ \nonumber
x_4 &=& \frac{\alpha y_{1} y_{3} + \frac{1}{\alpha} y_{2} y_{4}}{\sum y_{i}} y_{4} \\ \nonumber
|\frac{\partial (x_{i})}{\partial (y_{i})}| &=& 2 (\frac{\alpha y_{1} y_{3} + \frac{1}{\alpha} y_{2} y_{4}}{\sum y_{i}})^4
\end{eqnarray}

\begin{eqnarray}
I &=& \frac{2}{(- s u)^{2 - \frac{D}{4}}} \int_{0}^{\infty} \prod d y_{i} (\frac{\alpha y_{1} y_{3} + \frac{1}{\alpha} y_{2} y_{4}}{\sum y_{i}})^4 \frac{\delta (1 - \alpha y_{1} y_{3} - \frac{1}{\alpha} y_{2} y_{4})}{(\alpha x_{1} x_{3} + \frac{1}{\alpha} x_{2} x_{4})^{4-\frac{D}{2}}}  \\ \nonumber
&=& \frac{2}{(- s u)^{2 - \frac{D}{4}}} \int_{0}^{\infty} \prod d y_{i} (\sum y_{i})^{4 - D} \delta (1 - \alpha y_{1} y_{3} - \frac{1}{\alpha} y_{2} y_{4}) \\ \nonumber
\end{eqnarray}
and upon  the additional change of variables
\begin{eqnarray}
y_1 &=& \frac{1}{\sqrt{\alpha}} ( \tilde{y}_{1} - \tilde{y}_{3}) \quad y_3 = \frac{1}{\sqrt{\alpha}} ( \tilde{y}_{1} + \tilde{y}_{3})\\ \nonumber
y_2 &=& \sqrt{\alpha} ( \tilde{y}_{2} - \tilde{y}_{4}) \quad y_4 = \sqrt{\alpha} ( \tilde{y}_{2} + \tilde{y}_{4}) \\ \nonumber
\end{eqnarray}
we obtain
\begin{eqnarray}
\delta (1 - \alpha y_{1} y_{3} - \frac{1}{\alpha} y_{2} y_{4}) \to \delta (1 - [y_{1}^2 + y_{2}^2 - y_{3}^2 - y_{4}^2] )
\end{eqnarray}
It is clear  that we find  ourselves with the integration of some function over the region in $AdS_{3}$.

\section{Low-energy amplitudes  in the Higgsed phase}

Let us comment on the appearance of the similar $AdS_3$ geometry in the Higgsed phase
in the low-energy scattering regime. To this aim  note that the one loop
low-energy amplitude can be derived from the one-loop effective action in the
external field $F^{a}_{\mu\nu}$. It is assumed that the theory is in the Higgsed phase
and the invariants of the external field obey the relation $F^2/m^4<<1$,  $F\tilde{F}/m^4<<1$,
where $m$ is the mass of the particle in the loop.

The low-energy amplitude can be obtained by expanding  the effective action in the
proper power of the external field and the type of the amplitude is fixed by the
chiral structure of the external field. In particular the effective action
in the external self-dual field yields the generating function for the
all-plus(or all-minus) amplitudes which are known to vanish in SUSY case.
To get the $N^{k}MHV$ amplitude one has to expand the effective action in
$k+2$ power of $F_{-}$ and any number of $F_{+}$.

The explicit expression for the effective action in ${\cal N}=4$ $SU(2)$ SYM in the abelian background $F_{\mu \nu} = \bar{F}_{\mu \nu} \frac{\sigma_{3}}{2}$ with non-zero
scalar VEV $m$ reads as \cite{Buchbinder:1999jn}
\beq
{\cal L}_{eff}=- \frac{1}{4 \pi^2}\int^{\infty}_{0} \frac{ds}{s^3} e^{- m^2 s} \frac{f_1 f_2 s^2}{\sinh(f_1 s) \sinh(f_2 s)}
(\cosh(f_1s) - \cosh(f_2 s))^2
\label{act}
\eeq
where $f_1$ and $f_2$ are Euclidean field invariants; $m$ which can be written through proper invariants of external field
is the scalar VEV.
While going to Minkowski space $F_{0 i}^{E} \to i F_{0 i}^{M}$,
$F^2_{E} \to - F^2_{M}$, $F_{E} \tilde{F}_{E} \to  i F_{M} \tilde{F}_{M}$.
Then
\begin{eqnarray}
f_{1}^2 &\to& b^2 \\
f_{2}^2 &\to& - a^2
\end{eqnarray}
(here we adopt $(a,b)$ notations used in \cite{shub}).

In Minkowski space it is assumed that the external field is the sum of the plane waves

\beq
F_{tot}=\sum_i F_i \qquad F^{\pm}_{i,\mu\nu}= k_{i,\mu}e^{\pm}_{i\nu} -
 k_{i,\nu}e^{\pm}_{i\mu}
\eeq

To extract
the expression for the low energy amplitudes let us expand similar to \cite{shub}
the invariants in chiral components
\begin{eqnarray}
\frac{F_{tot}F_{tot}}{4} = \chi_{+} +\chi_{-}, \quad \frac{F_{tot}\tilde{F}_{tot}}{4}=-i(\chi_{+} - \chi_{-}) \\ \nonumber
\end{eqnarray}
where we have used
\beq
\chi_{+}=\frac{1}{2} \sum_{1\le i<j\le N} [i j]^2 \
\eeq
\beq
\chi_{-}=\frac{1}{2} \sum_{1\le i<j\le N} \langle i j \rangle^2
\eeq
and the standard spinor helicity notations are implied.
In terms of these variables the invariants involved into the effective action reads as
\beq
a=\sqrt{\chi_{+}} + \sqrt{\chi_{-}} \qquad b= - i (\sqrt{\chi_{+}} - \sqrt{\chi_{-}})
\eeq
\beq
f_{1}=\sqrt{\chi_{+}} + \sqrt{\chi_{-}} \qquad f_{2}= \sqrt{\chi_{+}} - \sqrt{\chi_{-}}
\eeq
\beq
{\cal L}_{eff} = - \frac{m^4}{4 \pi^2}\sum_{N=4}^{\infty}
 \frac{1}{(m^2)^{N}} \sum_{{k=0}\atop{k\,{\rm even}}}^{N}
\,c_{{\cal N} = 4} (\frac{k}{2},\frac{N-k}{2}) \chi_+^{k\over 2}\chi_-^{N-k\over 2}
\eeq
From this expression one can extract $N^{k-2}MHV$ low-energy $N$-particle amplitudes.
Explicit expression for the amplitude is given by $c_{{\cal N}=4} (\frac{k}{2},\frac{N-k}{2})$
multiplied by some kinematical factor (see \cite{shub}).

Here we find special properties of the low-energy amplitudes (for explicit formulas see appendix C):
\begin{itemize}
\item the effective action vanishes at the self-dual points $f_1=f_2$
in agreement with the vanishing of all-plus(minus) amplitudes in ${\cal{N}}=4$ SYM;
\item amplitudes with odd number of particles are zero;
\item amplitudes with odd number of positive (negative) helicity particles are zero (particularly NMHV amplitudes);
\item the only non-zero MHV amplitude is four-particle one.
\end{itemize}

During calculation we observed the following identities, which lie in the heart of the nullification of the higher points MHV amplitudes:
\begin{eqnarray}
\sum_{l=0}^{N} l^{\alpha} \frac{(1 - 2^{2l - 1} - 2^{2(N-l)-1}) B_{2l} B_{2(N-l)}}{\Gamma(2 l +1) \Gamma(2(N-l)+1)} \\ \nonumber
= - \frac{1}{4} \sum_{l=0}^{N} l^{\alpha} \left(\frac{(1 - 2^{2l - 1}) B_{2l}}{\Gamma(2 l +1) \Gamma(2(N-l))} + \frac{(1 - 2^{2(N-l)-1}) B_{2(N-l)}}{\Gamma(2 l) \Gamma(2(N-l)+1)} \right)
\end{eqnarray}
which we found to be true for $N \geq 3$ and $0 \leq \alpha \leq 3$.
These convolution identities are in the spirit of the ones that were considered in \cite{Dunne:2004uk}, but we have not found the same ones.
It is very natural to expect that supersymmetric effective actions serve as another rich source of
convolution identities for Bernoulli numbers.

Note that somewhat similar one-loop  nullification phenomena has been discovered
for the  amplitudes in QED \cite{mahlon} where it was found that
only non-vanishing all-plus(minus) amplitudes involves four
external legs.
It is natural to assume that some hidden symmetry is behind the
nullification phenomena of the low-energy amplitude. The
candidate in the ${\cal N}=4$ case is the Yangian symmetry and the possible arguments
implying the nullification could be similar to ones applied for the
nullification of the threshold amplitudes in \cite{gs}.

Now let us turn to the comment on the underlying $AdS$ geometry following \cite{gorly}.
In that paper it was remarked that the one-loop effective action
admits the geometrical interpretation. Namely, one can interpret the integrand
in the integral over the proper time as  transition amplitude for the
particle of some mass depending on the space-time spin in $AdS_3$. The proper time
measures the length of the corresponding geodesics. Even more suitable interpretation
emerges if we consider the $AdS_2$ geometry and the proper time $s$ measures the
length of the ``Wilson loop'' area. The low-energy effective action (\ref{act}) acquires the form of
the matrix element of some operator in the $2d$ gravity
\beq
{\cal L}_{eff}\propto \int ds \Psi_1(s)O\Psi_2(s)
\eeq
where $\Psi(s)$ is the "wave function" in $2d$ gravity and $O$ is some function of $s$. It can
be a little bit schematically written in the "length representation" of the wave functions $|f \rangle$
\beq
{\cal L}_{eff}(f_1,f_2)\propto \langle f_1|(\overleftarrow{L} - \overrightarrow{L})^2|f_2 \rangle
\eeq
where $L$ is the length operator acting on the corresponding state. In this representation
the Teichm\"{u}ller  phase space is implied and the momenta of external gluons entering
the invariants of the external fields mark the corresponding wave functions.

It is useful to have in mind the first-quantized representation of the effective action as the sum over the paths $C$
\beq
{\cal L}_{eff}(f_1,f_2)= \sum_{C}\exp(-i m L(C))\exp(i\Phi(C)) \langle W(C) \rangle
\eeq
where $L(C)$ is the length and $\Phi$ is the spin factor. That is the
low-energy limit of the amplitudes in the Higgs phase corresponds
to a kind of Fourier transform of the Wilson loops with respect
to the length.
Remark also that the low-energy effective action in the external field  can be used to calculate
some BPS invariants in the spirit  of  \cite{vafa}. It could be interesting to recognize the corresponding invariants which are related to the low-energy $N^k$MHV amplitudes. We hope to discuss these issues
elsewhere.

\section{Conclusion}
In this paper we have looked for the generalization of the amplitude-Wilson polygon
duality for the cases when the massive particles are at the external or internal legs
in the box diagrams.
We have suggested a version of  duality
for off-shell external legs and amplitudes at one-loop level. It implies
some modification of the dual object and instead of the Wilson
polygon the Wilson tetrahedrons emerge. The duality derivation is similar to the
on-shell case and involves the particular change of variables.  Note that
our result is the useful step toward the dual description of the NMHV amplitudes
which involve 2mh,3mh and 4mh boxes. However we do not know how the
coefficients in front of the boxes can be derived in the geometric manner.

One-loop duality appears to be quite transparent however its two-loop generalization
deserves the formulation of the proper regularization of the "Wilson polygon".
The naive $AdS$ inspired regularization does not work and it would be  important
to find the regularization distinct from dimensional one. It it also very
interesting to make the link of our approach with the twistor picture.

We have made also a few comments concerning the appearance of the $AdS$-like geometry
in the one-loop calculation. It turns out that the box diagram can be attributed to the
integration of particular function over the $AdS_3$. In the deep Higgsed regime
we have found the interesting nullification phenomena for the
low-energy amplitudes. It would be interesting to develop
the duality arguments which would exchange small mass and large mass limits.

We are grateful to G. Korchemsky for the useful discussions.  (A.G.) thanks
CPhT at Sacley and SUBATECH, where the part of the work has been done for the hospitality.
The work was supported by grant PICS- 07-0292165 (A.G), RFBR-09-02-00308 (A.G.) and
CRDF -  RUP2-2961-MO-09 (A.G.)

\section*{Appendix A \qquad Basic notations and integrals}
\begin{eqnarray}
a &=& \frac{g^2 N}{8 \pi^2} \\  \nonumber
I^{N}(D;{\nu_{k}}) &=& - i \pi^{-\frac{D}{2}} (\mu_{IR}^2)^{- \epsilon} \int d^{D}l \frac{1}{A^{\nu_{1}}_{1}A^{\nu_{2}}_{2}...A^{\nu_{N}}_{N}}  \nonumber
\end{eqnarray}
$$\begin{array}{cc}
C_{F} = \frac{N^2 - 1}{2 N}, & c_{\Gamma} = (\mu_{IR}^2)^{\epsilon} \frac{\Gamma(1 - \epsilon) \Gamma(1 + \epsilon)^2}{\Gamma(1 + 2 \epsilon)}, \\
C_{A} = N, & (\mu_{IR}^2)^{- \epsilon} = (\pi \mu_{UV}^2)^{\epsilon}. \\
\end{array}$$

\begin{eqnarray}
I_{3}(m^2,0,0|4+2 \epsilon) &=& - \frac{c_{\Gamma}}{\epsilon^2} (- m^2)^{-1+\epsilon} \\
I_{3}(m^2_{1},m^2_{2},0|4+2 \epsilon) &=& - \frac{c_{\Gamma}}{\epsilon^2} \frac{(-m_{1}^2)^{\epsilon}-(-m_{2}^2)^{\epsilon}}{(-m_{1}^2)-(-m_{2}^2)}
\end{eqnarray}

For the Wilson cusp diagram  with $m^2 =2 (p_{i}, p_{j})$:
\begin{eqnarray}
G^{F}_{\mu \nu}(x-y) &=& - \eta_{\mu \nu} \frac{ (\pi \mu_{UV}^2)^{\epsilon}}{4 \pi^2} \frac{\Gamma(1 - \epsilon)}{(-(x-y)^2 + i \epsilon)^{1 - \epsilon}}\\
\frac{I^{cusp}(m^2|4 - 2 \epsilon)}{(p_{i}, p_{j})} &=& a (- m^2)^{-1+\epsilon} \frac{(\pi \mu_{UV}^2)^{\epsilon} \Gamma(1 - \epsilon)}{\epsilon^2}
\end{eqnarray}

\section*{Appendix B \qquad Connection of scalar integrals in different dimensions}

Here we briefly explain the connection between the scalar integrals in different dimensions \cite{Nizic2}.
Suppose, we have the following scalar integral (see fig.9)
\begin{eqnarray} \label{conn1}
I^{N}(D;{\nu_{k}}) \equiv - i \pi^{-\frac{D}{2}} (\mu^2)^{\epsilon} \int d^{D}l \frac{1}{A^{\nu_{1}}_{1}A^{\nu_{2}}_{2}...A^{\nu_{N}}_{N}} \\ \nonumber
\end{eqnarray}

\begin{figure}
\begin{center}
\fcolorbox{white}{white}{
  \begin{picture}(204,174) (68,-79)
    \SetWidth{1.0}
    \SetColor{Black}
    \Arc(167,9)(53.009,1,361)
    \Line(99,-55)(129,-29)
    \Line(74,5)(114,6)
    \Line(204,47)(233,75)
    \Line(220,6)(255,7)
    \Line(168,-44)(168,-78)
    \Line(167,62)(167,94)
    \Line(204,-29)(234,-58)
    \Line(127,45)(97,74)
    \Text(88,56)[lb]{\Black{$p_{3}$}}
    \Text(176,-79)[lb]{\Black{$p_{N}$}}
    \Text(237,-48)[lb]{\Black{$p_{N-1}$}}
    \Text(184,-55)[lb]{\normalsize{\Black{$l+r_{N}$}}}
    \Text(131,-53)[lb]{\normalsize{\Black{$l+r_{1}$}}}
    \Text(89,-19)[lb]{\normalsize{\Black{$l+r_{2}$}}}
    \Text(89,22)[lb]{\normalsize{\Black{$l+r_{3}$}}}
    \Text(217,-23)[lb]{\normalsize{\Black{$l+r_{N-1}$}}}
    \Text(101,-66)[lb]{\Black{$p_{1}$}}
    \Text(65,-13)[lb]{\Black{$p_{2}$}}
  \end{picture}
}
\end{center}
\caption{General one-loop diagram}
\end{figure}
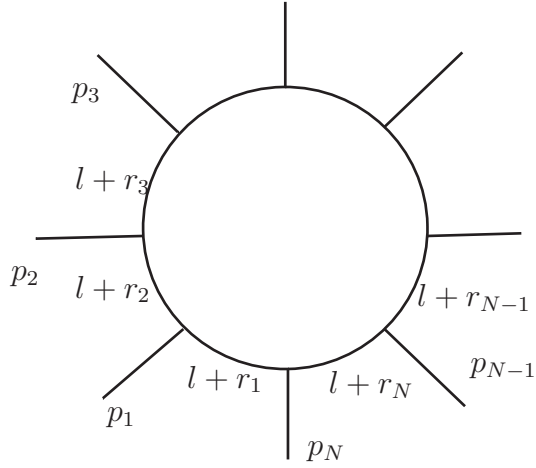

then it can be shown that
\begin{eqnarray} \label{conn}
I^{N} (D-2;{\nu_{k}}) = \sum_{i=1}^{N} z_{i} I^{N} (D-2;{\nu_{k} - \delta_{k i}}) \\ \nonumber
+ (D - 1 - \sum_{j=1}^{N} \nu_{j}) z_{0} I^{N} (D;{\nu_{k}})
\end{eqnarray}
where
\begin{eqnarray} \label{conn}
\sum_{i=1}^{N} (r_{i} - r_{j})^2 z_{i} &=& 1 \\ \nonumber
z_{0} &=& \sum_{i=1}^{N} z_{i}
\end{eqnarray}
In the main body of the text we choose $D=6 + 2 \epsilon$, $N=4$, $\nu_{i}=1$.

We are interested in the case of $D=6$ two-mass easy boxes and their connection with $D=4$ ones \cite{Nizic2} hence
\begin{eqnarray}\nonumber
I^{2me}(6 + 2 \epsilon) = \frac{1}{(1 + 2 \epsilon) z_{0}} (I^{2me}(4 + 2 \epsilon) - \sum_{i=1}^{4} z_{i} I^{2me}(4 + 2 \epsilon; 1 - \delta_{k i} ) )
\end{eqnarray}
where
\begin{eqnarray}\nonumber
z_{0}= \sum_{i=1}^{4} z_{i} &=& 2 \frac{s + u - m_{2}^2 - m_{4}^{2}}{s u - m^2_{2} m^2_{4}} \\ \nonumber
z_{1} &=& \frac{u-m_{2}^{2}}{s u - m^2_{2} m^2_{4}} \\ \nonumber
z_{2} &=& \frac{s-m_{4}^{2}}{s u - m^2_{2} m^2_{4}} \\ \nonumber
z_{3} &=& \frac{u-m_{4}^{2}}{s u - m^2_{2} m^2_{4}} \\ \nonumber
z_{4} &=& \frac{s-m_{2}^{2}}{s u - m^2_{2} m^2_{4}} \\ \nonumber
\end{eqnarray}
 It can be easily seen the $\sum_{i=1}^{4} z_{i} I^{4}(4 + 2 \epsilon; 1 - \delta_{k i})$ does precisely the job of taking  the finite part.

\section*{Appendix C \qquad Low-energy amplitudes in ${\cal N}=4$ SYM}

\begin{eqnarray}
{\cal L}_{eff}=- \frac{1}{4 \pi^2}\int^{\infty}_{0} \frac{ds}{s^3} e^{- m^2 s} \frac{f_1 f_2 s^2}{\sinh(f_1 s) \sinh(f_2 s)}
(\cosh(f_1s) - \cosh(f_2 s))^2
\label{act}
\end{eqnarray}
$$x = f_{1} s$$ $$y = f_{2} s$$

\begin{eqnarray}
\frac{f_1 f_2 s^2}{\sinh(f_1 s) \sinh(f_2 s)}
(\cosh(f_1s) - \cosh(f_2 s))^2 =\\ \nonumber
2 (\frac{x}{\sinh x} \frac{y}{\sinh y} - \frac{x}{\tanh x} \frac{y}{ \tanh y}) + x \sinh x \frac{y}{\sinh y} + \frac{x}{\sinh x} y \sinh y
\end{eqnarray}

\begin{eqnarray}
\frac{x}{\tanh x} = \sum_{n=0}^{\infty} \frac{2^{2n} B_{2n}}{2n!} x^{2n}
\end{eqnarray}
\begin{eqnarray}
\frac{x}{\sinh x} = \sum_{n=0}^{\infty} \frac{2(1-2^{2n-1})B_{2n}}{2n!} x^{2n}
\end{eqnarray}
\begin{eqnarray}
x \sinh x = \sum_{n=0}^{\infty} \frac{x^{2(n+1)}}{(2n+1)!}
\end{eqnarray}

\begin{eqnarray} \label{action}
{\cal L}_{eff}=- \frac{m^4}{4 \pi^2} \sum_{n=0}^{\infty} \sum_{l=0}^{\infty} \frac{\Gamma(2 (l+n-1))}{(m^2)^{2(l+n)}} f_{1}^{2 n} f_{2}^{2 l} {\cal C}(n,l) \\
=- \frac{m^4}{8 \pi^2} \sum_{n,l=0}^{\infty}  \frac{\Gamma(2 (l+n-1))}{(m^2)^{2(l+n)}} (f_{1}^{2 n} f_{2}^{2 l} + f_{2}^{2 n} f_{1}^{2 l}){\cal C}(n,l)
\end{eqnarray}
\begin{eqnarray}
{\cal C}(n,l) &=& {\cal C}_{1}(n,l) + {\cal C}_{2}(n,l) + {\cal C}_{2}(l,n) \\
{\cal C}_{1}(n,l) &=& 8 \frac{B_{2 n} B_{2 l}}{(2 n)! (2 l)!} (1 - 2^{2 n -1} - 2^{2 l -1}) \\
{\cal C}_{2}(n,l) &=& 2 \frac{B_{2 l}}{(2 n - 1)! (2 l)!} (1 - 2^{2 l -1}), \quad n \geq 1 \\
{\cal C}_{2}(0,l) &=& 0
\end{eqnarray}
Due to symmetries of original expression ${\cal C}(n,l) = {\cal C}(l,n)$

\begin{eqnarray}
f_{1}=\sqrt{\chi_{+}} + \sqrt{\chi_{-}} \qquad f_{2}= \sqrt{\chi_{+}} - \sqrt{\chi_{-}}
\end{eqnarray}

\begin{eqnarray}
f_{1}^{2 n} f_{2}^{2 l} + f_{2}^{2 n} f_{1}^{2 l} = \sum_{\alpha=0}^{2n} \sum_{\beta=0}^{2l} \sqrt{\chi_{+}}^{\alpha + \beta} \sqrt{\chi_{-}}^{2(n+l)- (\alpha + \beta)} C^{2n}_{\alpha} C^{2l}_{\beta} ( (-1)^{\alpha} + (-1)^{\beta})
\end{eqnarray}
$$ \alpha + \beta = 2 K $$

\begin{eqnarray}
f_{1}^{2 n} f_{2}^{2 l} + f_{2}^{2 n} f_{1}^{2 l} &=&  \sum_{K=0}^{n+l}  \chi_{+}^{K} \chi_{-}^{(n+l)- K} \Upsilon(n,l,K) \\
\Upsilon(n,l,K) &=& 2 \sum_{\beta=0}^{2l}  C^{2n}_{2 K - \beta} C^{2l}_{\beta} (-1)^{\beta} \Theta (\beta > 2 K - 2 n) \Theta (\beta < 2 K)
\end{eqnarray}
Let's return to (\ref{action}) and make change $N = n+l$
\begin{eqnarray} \label{action}
{\cal L}_{eff} &=&- \frac{m^4}{4 \pi^2} \sum_{n=0}^{\infty} \sum_{l=0}^{\infty} \frac{\Gamma(2 (l+n-1))}{(m^2)^{2(l+n)}} f_{1}^{2 n} f_{2}^{2 l} {\cal C}(n,l) \\
&=&- \frac{m^4}{4 \pi^2} \sum_{N=1}^{\infty}  \frac{(2 N -3)!}{(m^2)^{2 N}} \sum_{K=0}^{N} \chi_{+}^{K} \chi_{-}^{N - K} {\cal A}_{K,N} \\
{\cal A}_{K,N} &=& \sum_{l=0}^{N} {\cal C}(N-l,l) \Upsilon(N-l,l,K)
\end{eqnarray}

All plus and all minus amplitudes are zero, because
\begin{eqnarray}
\Upsilon(N-l,l,0) = 2 \\
\sum_{l=0}^{N} {\cal C}(N-l,l) = 0
\end{eqnarray}

In fact non-zero amplitudes appear when $N \geq 2$ which corresponds to four particles.

$2 K$ - number of particles with positive helicity.

$2 N$ - total number of particles.

Using the notation in the main body of the text
\begin{eqnarray}
c_{{\cal N} = 4} (\frac{k}{2},\frac{N-k}{2}) = {\cal A}_{K,N}
\end{eqnarray}

\section*{Appendix D \qquad Geometry of divergencies}
If we have off-shell regularization then we are free to
think about triangle or  about the box, since they are defined via
the same functions - the only difference is hidden in the "glued" kinematics.

Let's introduce standard notations

\beq
x=\frac{m_{1}^2 m_{3}^2}{s u} \qquad
y=\frac{m_{2}^2 m_{4}^2}{s u}
\eeq

\begin{eqnarray}
\lambda (x, y) = \sqrt{(1 - x - y)^2 -4 x y} \\ \nonumber
\rho (x, y) = \frac{2}{1 - x - y - \lambda}
\end{eqnarray}
and
\begin{eqnarray}
\Phi (x , y) = \frac{1}{\lambda} \{ 2 [ Li_{2}(- \rho x) + Li_{2}(- \rho y)] + \log \frac{y}{x} \log \frac{1 + \rho y}{1 + \rho x} + \log (\rho x) \log (\rho y) + \frac{\pi^2}{3}  \}
\end{eqnarray}
The massive box diagram is given by
\begin{eqnarray}
I_{box} = \frac{i \pi^2}{ s u } \Phi (x , y)
\end{eqnarray}
For the triangle we obtain
\begin{eqnarray}
x = \frac{p_1^2}{p_3^2}, \quad y = \frac{p_2^2}{p_3^2} \nonumber
\end{eqnarray}
and the massive triangle diagram is  given by
\begin{eqnarray}
I_{\triangle} = \frac{i \pi^2}{p_{3}^2} \Phi (x , y)
\end{eqnarray}
These diagrams can be interpreted in geometrical terms \cite{dav}.

Then dihedral angles of ideal tetrahedron are
\beq
cos \psi_{12}= \frac{1 + y - x}{2 \sqrt{y}}
\eeq
\beq
cos \psi_{13}= \frac{x + y - 1}{2 \sqrt{x y}}
\eeq
\beq
cos \psi_{14}= \frac{1 + x - y}{2 \sqrt{x}}
\eeq

\beq
\mbox{Cl}_2(x)=\mbox{Im} [\mbox{Li}_2(e^{ix})]= - \int_{0}^{x} dy ~ \mbox{ln}|2 \sin y/2| = - \sin \theta \int_{0}^{1} \frac{d z \log z}{1 - 2 z \cos \theta + z^2}
\eeq

\beq
2i\Omega^{(4)}= \mbox{Cl}_2(2 \psi_{12}) +\mbox{Cl}_2(2 \psi_{13}) +\mbox{Cl}_{2}(2 \psi_{23}) = \lambda(x,y,1) \Phi_{Dav} (x,y)
\eeq
where the K\"{a}llen function $\lambda(x,y,z)$ is defined as

\beq
\lambda(x,y,z)=x^2 +y^2 +z^2 -2xy -2yz -2zx
\eeq

Going on-shell means that $ x \to 0$ or $y \to 0$ and this limit should
be understood in terms of the analytical continuation.

\subsection*{Geometrical regions}

We can start with the kinematical region where geometrical picture is clear and
\beq
\psi_{12}+ \psi_{13} + \psi_{14} = \pi
\eeq
Generically one can  consider simple
triangle with these angles instead of the  hard ideal hyperbolic tetrahedron.
This geometrical picture is valid when $|\cos \psi|<1$ or $(\sqrt{y}-1)^2 < x < (\sqrt{y}+1)^2$.
At the boundary $x=(\sqrt{y} \pm 1)^2$ the volume vanishes and equivalently
area of the triangle vanishes as well.
Thus,  to go on-shell we need to make the proper analytic continuation.

\subsection*{Analytic continuation of basic functions}
Let us introduce notation
\beq
\mbox{Im} [f(z)] = \frac{f(z)-f(z^{*})}{2 i} \rightarrow \mbox{Odd}[f(z)] = \frac{f(z) - f(-z)}{2}
\eeq
Then
\beq
\mbox{Clh}_2(x)=\mbox{Odd} [\mbox{Li}_2(e^{x})]= - \int_{0}^{x} dy ~ \mbox{ln}|2 \sinh y/2| = - \sinh x \int_{0}^{1} \frac{d z \log z}{1 - 2 z \cosh x + z^2}
\eeq
\beq
0 < \mbox{arccosh} ~ x < \infty
\eeq

\begin{itemize}
\item $x>(\sqrt{y}+1)^2$
\end{itemize}
Note appearance of the sign in $\psi_{12}$

\beq
\cosh -\psi_{12}=- \frac{1 + y - x}{2 \sqrt{y}}
\eeq
\beq
\cosh \psi_{13}= \frac{x + y - 1}{2 \sqrt{x y}}
\eeq
\beq
\cosh \psi_{14}= \frac{1 + x - y}{2 \sqrt{x}}
\eeq
and the sum vanishes
\beq
\psi_{12}+ \psi_{13} + \psi_{14} = 0
\eeq

\beq
2 \Omega^{(4)}= \mbox{Clh}_2(2 \psi_{12}) +\mbox{Clh}_2(2 \psi_{13}) +\mbox{Clh}_{2}(2 \psi_{23})
\eeq
\begin{itemize}
\item $x < (\sqrt{y}-1)^2$
\end{itemize}

\beq
\cosh \psi_{12}=\frac{1 + y - x}{2 \sqrt{y}}
\eeq
\beq
\cosh \psi_{13}= \frac{x + y - 1}{2 \sqrt{x y}}
\eeq
\beq
\cosh -\psi_{14}=- \frac{1 + x - y}{2 \sqrt{x}}
\eeq

\beq
\psi_{12}+ \psi_{13} + \psi_{14} = 0
\eeq

\beq
2 \Omega^{(4)}= \mbox{Clh}_2(2 \psi_{12}) +\mbox{Clh}_2(2 \psi_{13}) +\mbox{Clh}_{2}(2 \psi_{23})
\eeq
Massless limit $x \rightarrow 0$ corresponds to $\psi_{13} \rightarrow \infty$ and $\psi_{14} \to - \infty$

\end{document}